\documentclass{article}
\usepackage[utf8]{inputenc}
\usepackage{graphicx}
\usepackage{booktabs} 
\usepackage{algpseudocode}
\usepackage{algorithm2e}
\usepackage{float}
\newfloat{algorithm}{t}{lop}
\usepackage{eqparbox}
\usepackage{multicol}
\usepackage{booktabs}
\usepackage{url}
\usepackage{color}
\usepackage{multirow}
\usepackage{mathtools}
\usepackage{amsmath}
\usepackage{amssymb}

\begin{document}

\thispagestyle{empty}
\title{Estimating adaptive cruise control model parameters from on-board radar units}

\author{\\
{\large Yanbing Wang}\vspace{-1ex}\\
{\normalsize Department of Civil and Environmental Engineering}\vspace{-1ex}\\
{\normalsize Institute for Software Integrated Systems}\vspace{-1ex}\\
{\normalsize Vanderbilt University }\vspace{-1ex}\\
{\normalsize 1025 16th Avenue South}\vspace{-1ex}\\
{\normalsize Nashville, TN 37212 }\vspace{-1ex}\\
{\normalsize yanbing.wang@vanderbilt.edu}\\
{\normalsize{} }\\
{\large George Gunter (corresponding author)}\vspace{-1ex}\\
{\normalsize Department of Civil and Environmental Engineering}\vspace{-1ex}\\
{\normalsize Institute for Software Integrated Systems}\vspace{-1ex}\\
{\normalsize Vanderbilt University }\vspace{-1ex}\\
{\normalsize 1025 16th Avenue South}\vspace{-1ex}\\
{\normalsize Nashville, TN 37212 }\vspace{-1ex}\\
{\normalsize george.l.gunter@vanderbilt.edu }\\
{\normalsize{} }\\
{\large Matthew Nice}\vspace{-1ex}\\
{\normalsize Department of Electrical Engineering and Computer Science}\vspace{-1ex}\\
{\normalsize Institute for Software Integrated Systems}\vspace{-1ex}\\
{\normalsize Vanderbilt University }\vspace{-1ex}\\
{\normalsize 1025 16th Avenue South}\vspace{-1ex}\\
{\normalsize Nashville, TN 37212 }\vspace{-1ex}\\
{\normalsize matthew.nice@vanderbilt.edu}\\
{\normalsize{} }\\
{\large Daniel B. Work}\vspace{-1ex}\\
{\normalsize Department of Civil and Environmental Engineering}\vspace{-1ex}\\
{\normalsize Institute for Software Integrated Systems}\vspace{-1ex}\\
{\normalsize Vanderbilt University }\vspace{-1ex}\\
{\normalsize 1025 16th Avenue South}\vspace{-1ex}\\
{\normalsize Nashville, TN 37212 }\vspace{-1ex}\\
{\normalsize{} }}

\maketitle

\newpage

\thispagestyle{empty}
\section*{Abstract}
Two new methods are presented for estimating car-following model parameters using data collected from \textit{Adaptive Cruise Control} (ACC) enabled vehicles. The vehicle is assumed to follow a constant time headway relative velocity model in which the parameters are unknown and to be determined. The first technique is a batch method that uses a least-squares approach to estimate the parameters from time series data of the vehicle speed, space gap, and relative velocity of a lead vehicle. The second method is an online approach that uses a particle filter to simultaneously estimate both the state of the system and the model parameters. Numerical experiments demonstrate the accuracy and computational performance of the methods relative to a commonly used simulation-based optimization approach. The methods are also assessed on empirical data collected from a 2019 model year ACC vehicle driven in a highway environment. Speed, space gap, and relative velocity data are recorded directly from the factory-installed radar unit via the vehicle's CAN bus. All three methods return similar mean absolute error values in speed and spacing compared to the recorded data. The least-squares method has the fastest run-time performance, and is up to 3 orders of magnitude faster than other methods. The particle filter is faster than real-time, and therefore is suitable in streaming applications in which the datasets can grow arbitrarily large.


\section{Introduction}\label{sec:Intro}
The rising penetration rate of \textit{Society of Automotive Engineers} (SAE) level one and level two vehicles on roadways around the world is creating new traffic flows that are a combination of human drivers and vehicle automation systems. SAE level one and level two vehicles assist with some of the driving tasks such as car following and lane keeping. For example, one common feature that is now available on many vehicles is \textit{adaptive cruise control} (ACC). Like human driven vehicles, it is important to have simple and accurate models of the automated vehicles so that their impacts on the total traffic flow can be determined. To this end, several works~\cite{gunter2019are,gunter2019arxiv,milanes2014modeling,knoop2019platoon} are exploring quantitatively and qualitatively properties of commercially available vehicles, such as their string stability.

Motivated by this line of work, this article considers the problem of estimating the parameters of a car following model to characterize the behavior of an automated vehicle. Car following models were originally introduced to describe the behavior of a human driver and are typically written as an ordinary differential equation or difference equation. Famous examples include the Intelligent Driver Model, Gipps Model, and several variants of the Optimal Velocity Model \cite{kesting2010enhanced,gipps1981behavioural,BandoHesebeNakayama1995}.  These models output a driver's instantaneous acceleration ($\dot{v}$) as a function of their current speed ($v$), inter-vehicle spacing ($s$), and/or relative speed difference to a vehicle in front ($\Delta v$). Each of these models have additional parameters that can take on ranges of different values to describe different driving behaviors. Common examples include the minimum inter-vehicle spacing a driver is willing to tolerate, or a maximum speed a driver is willing to drive at. 

While in some cases it is reasonable to simply assume model parameter values based on assumptions about driving behavior, a more rigorous technique for selecting these values is to calibrate a proposed car-following model against recorded vehicle trajectories, a famous example of such trajectories being the \textit{Next Generation SIMulation} (NGSIM) dataset. With the increasing proliferation of GPS units on vehicles and mobile devices, and the rise in sensing accuracy, more datasets of this nature are expected to become available~\cite{MobileCentury}. As such, it is important to develop fast and accurate estimation routines which can learn new models from driving behavior, particularly as automated vehicles enter into the traffic stream alongside human drivers.

Several works have looked at the car-following estimation problem in the context of human drivers, with several in the context of the above mentioned NGSim dataset~\cite{NGSimCalibration1,NGSimCalibration2,NGSimCalibration3}. Kesting and Trieber proposed a methodology to estimate parameters for the Intelligent Driver Model by minimizing the error between simulated driving trajectories and measure, via the use of a genetic search algorithm in order to account for non-convexity in the search space~\cite{KestingModelCalibration2008}. Punzo and Simonelli estimated optimal parameters for several different car-following models on a 4 vehicle platoon equipped with GPS tracking units, and used a batch optimization method in which a gradient-based optimization technique was used starting from several different initial points, again in order to account for potential non-convexity~\cite{PunzoModelCalibration2005}. This method is in the same family of techniques used in this article as a benchmark to compare the newly presented methods. In general it should be noted that many of these techniques find optimal parameters of non-linear models, and as such need to employ optimization techniques that account for this difficulty. 

The methods used in~\cite{PunzoModelCalibration2005,gunter2019are,gunter2019arxiv,milanes2014modeling} are offline batch methods that are applied only after all data has been collected. Online methods, in contrast, sequentially estimate the parameters while moving in a single pass through the data. Many of these methods use filtering techniques from traditional control theory to combine model parameter estimation with state information. In contrast to offline methods, online estimators typically are scalable to arbitrarily large datasets and can potentially be used in real-time settings. Such techniques have been demonstrated to be able to identify driving models such as the IDM that accurately reproduce driving behavior~\cite{Monteil2015RealTimeEO}. However, challenges to online parameter estimation exist. Monteil and Bouroche point out that careful experimental design is necessary to enable the identifiability of some parameters depending on the car following dynamics~\cite{Monteil_identifiability}. In addition, noise characterization can be non-trivial when calibrating a filtering algorithm~\cite{Beckmann_tuning}. 

The main contribution of this paper is to propose two new parameter estimation techniques for a common car-following model used to model ACC equipped vehicles, and to benchmark their performance against an existing approach. Each method is demonstrated by performing system identification of an ACC enabled vehicle using data collected from its on-board radar system.  The first new method uses a batch least-squares approach, while the second method is an online method based on particle filtering.  The two proposed algorithms are additionally compared against a more commonly used batch-optimization method for run-time and model accuracy.

The three methods are first tested on a numerical experiment where the data is synthetically generated from a model in which the true parameters are known, in order to compare the accuracy of the recovered parameters. Next, each method is implemented on a real dataset, which is collected by gathering radar data from an ACC enabled vehicle's on-board CAN bus. The vehicle records space gap, velocity, and relative velocity measurements. A timeseries of this data is used to estimate the model parameters using the three methods. The methods are compared in terms of the runtime, and the accuracy of the model with the learned parameters. As an illustration of the importance of the parameter estimates to the broader context of mixed traffic flows, we also estimate the string stability of the ACC system using the estimated parameters. 

The least-squares method is found to have the fastest performance on the empirical dataset, but will not scale as well as the particle filter for longer datasets, due to the fact it is a batch method. All of the learned models are found to achieve similar and low errors, suggesting that all methods are viable estimation techniques. 

The remainder of the paper is organized as follows. The section, \textit{Parameter Estimation Methodology}, outlines each estimation method. The section, \textit{Estimation On Synthetic Data}, demonstrates the estimation routines on synthetically generated (simulated) data, in order to understand the ability for the method to reconstruct a given model under no uncertainty. The section, \textit{Estimation On Real Data}, presents both the experimental protocol for collecting ACC radar measurements, and the results of parameter estimation on those results. Finally, the conclusion section summarizes the the performance of each technique and the overall contribution.

\section{Parameter Estimation Methodology}\label{sec:CalMethodology}
In this section, we briefly review a common model assumed for ACC vehicle dynamics, and provide a criterion for determining  the string stability of the model. We then review a standard simulation-based optimization method to estimate the model parameters, before introducing the new methods based on least-squares and particle filtering respectively.  

\subsection{Model Description and String Stability}\label{sec:ModelDescription}
In order to complete the estimation problem a candidate model must be used. While most estimation routines have focused on fitting car-following models to human drivers, recent work has shown that ACC vehicles may also be modeled accurately. With the increasing interest in how vehicles with autonomous features will affect traffic flow patterns~\cite{AutomatedTraffic_1,AutomatedTraffic_2} several works have looked at modeling ACCs using car-following models \cite{milanes2014modeling,gunter2019arxiv,gunter2019are}. The most commonly used model is the \textit{constant-time headway relative-velocity} (CTH-RV) car-following model, and is used in this work due to the its linear nature (which allows for a global string-stability analysis) and previous demonstration of modeling ACC systems well~\cite{gunter2019model_comp}. The CTH-RV is written as:

\begin{equation}
\label{eq:CTH-FL}
\dot{v}(t) = f(\boldsymbol{\theta}, s(t), v(t), \Delta v(t)) = k_1(s(t) - \tau v(t)) + k_2(\Delta v(t)), 
\end{equation}

\noindent where $s$, $v$, and $\Delta v$ are the space gap, speed, and speed difference between the ACC vehicle and a leading vehicle. The vector of model parameters $\boldsymbol{\theta} = [k_1, k_2, \tau]^T$ control the gain on the constant time headway term and the relative velocity term respectively, and the parameter $\tau$ is the time gap at equilibrium.

One important property of a car following model is whether it amplifies or dissipates disturbances in the traffic flow. The string stability of a vehicle determines if a small perturbation to the state of a proceeding (lead) vehicle in a platoon will either amplify (string-unstable) or dissipate (string stable) in magnitude as it is propagated back in the traffic. String unstable car following dynamics give rise to so called phantom traffic jams, which seem to arise without an obvious cause but are in fact a feature of the car following behavior. Given a car following model $f(\cdot,\cdot,\cdot, \cdot)$ as in \eqref{eq:CTH-FL}, it is easy to check the string stability by computing partial derivatives of the model with respect to its inputs~\cite{WilsonWard2011}:
\begin{equation}\label{eq:StringStability_General}
    \lambda \coloneqq \dfrac{f_s}{f_v^3}\left(\dfrac{f_v^2}{2}-f_{\Delta v}f_v-f_s\right),
\end{equation}
where $f_s:=\frac{\partial f}{\partial s}$, $f_{\Delta v}:=\frac{\partial f}{\partial \Delta v}$, and $f_v:=\frac{\partial f}{\partial v}$ are the partial derivatives of the car-following model. If $\lambda<0$ then the model is string stable and if $\lambda>0$ it is string unstable. In the context of \eqref{eq:CTH-FL} this condition can be written explicitly in terms of $k_1,k_2,$ and $\tau$:
\begin{equation}\label{StringStability_CTHFTL}
    \lambda(k_1,k_2,\tau) = \frac{k_1}{-k_1^3\tau^3}\left[\frac{k_1^2\tau^2}{2} + k_1 k_2 \tau -k_1\right].
\end{equation}

It is worth noting that $\lambda$ only describes global string stability for linear models, while for non-linear models it describes only local string stability around an equilibrium. Since \eqref{eq:CTH-FL} is a linear model in the state variables $(s,v,\Delta v)$ its string stability globally holds. Each of the studies that have used this model found that that ACC vehicle driving behavior is string unstable~\cite{gunter2019arxiv,gunter2019are}.

\subsection{Offline Batch Optimization}
Here a batch optimization scheme is reviewed in which the car-following model appears as a constraint in the optimization problem. The problem can be directly solved as a simulation-based optimization problem using standard descent-based optimization routines.  Given a timeseries dataset of following vehicle speeds, spacing values, leading vehicle speeds, and an initial guess for the parameters, this method solves the ODE and compares the simulated trajectory with the measured data. The simulated trajectories are then assessed according to the performance measure specified in the objective function. New parameters are selected to improve the performance measure until a (possibly local) minimum is obtained. This approach is considered to be a well-known technique for car-following parameter estimation ~\cite{PunzoModelCalibration2005}, and is considered in this work as a benchmark for the performance of the two other estimation methods proposed.

In this work optimal parameter values are optimized for according to the \textit{root mean squared error} (RMSE) between simulated inter-vehicle space gap data and recorded space gap data. It is worth noting that other error metrics might also be reasonable (e.g., errors based on the absolute error, or based on velocity measures). The RMSE space gap error is used here because it was found to produce reasonable results in previous works ~\cite{PunzoModelCalibration2005,KestingModelCalibration2008}. The general form optimization scheme is written as:

\begin{equation}\label{eq:estimationRoutine}
\begin{array}{rl}
{{\text{minimize}}}: & \sqrt{\frac{1}{T}\int_0^T{({s}^\text{m}(t)-s(t))^2}dt}\\
\text{subject to:}
 & \dot{v}(t) = f(\boldsymbol{\theta},s, v, \Delta v)\\
 & \dot{s}(t) = v_{l}(t) - v(t),\\
\end{array}
\end{equation}

\noindent with possible additional constraints on the initial conditions, and bounds on the parameters. In~\eqref{eq:estimationRoutine}, $f(\boldsymbol{\theta},s, v, \Delta v)$ corresponds to the car-following model in \eqref{eq:CTH-FL}. The term $v_{l}(t)$ is the lead vehicle speed as a function of time and is assumed to be available from measured data. Similarly, $s(t)^\text{m}$ denotes the measured spacing, which is compared to the spacing predicted by the model in the objective function. Alternative objective functions based on speed differences, or a weighted combination of speed and spacing can also be considered. The total time of the dataset and simulation is $T$. Because the ODE appears as a constraint, a typical approach to solve the problem uses simulation based optimization.   

It is important to note that the problem is nonlinear in the decision variables (the state and model parameters), and depending on the form of the car following model, it may also be non-convex. To combat this potential problem the optimization problem can be solved many times, with each run starting from randomly selected different initial candidate parameter values, as in ~\cite{PunzoModelCalibration2005}. 

\subsection{Least-Squares Matrix Problem}\label{sec:LSTQ_Section}
We next introduce a new and simple approach to estimate the model parameters of a linear car following model when the speed, spacing, and relative velocity are all directly and accurately measured.
Unlike~\eqref{eq:estimationRoutine}, the least-squares method proposed here does not require multiple starting points or solving solving multiple ODEs within each optimization run, substantially reducing the runtime. We briefly derive the least-squares formulation for the ACC car following model~\eqref{eq:CTH-FL} and show that the optimal parameters can be computed by taking a matrix pseudo-inverse (denoted by $\dagger$).

First we rewrite the continuous time ODE \eqref{eq:CTH-FL} in discrete-time using a forward Euler step scheme:
\begin{equation}\label{eq:CTH-FTL-Discrete}
\begin{array}{lll}
     v_{k+1}& = & v_{k} +f(\boldsymbol{\theta}, s_{k}, v_{k}, \Delta v_{k})\Delta T  \\
     &=& v_{k} + (k_1(s_{k} - \tau v_{k})\Delta T + k_2(v_{l,k} - v_{k})\Delta T)  \\
     &= & (1-(k_1\tau +k_2)\Delta T)v_{k} + (k_1 \Delta T)s_{k} + (k_2 \Delta T)v_{l,k},
\end{array}
\end{equation}
where $\Delta T$ is the timestep. It is selected to correspond to the frequency at which the speed, spacing, and relative velocity data is measured (e.g., on the order of 1/10 of a second for some sensor platforms including the experiments presented later in this work).

Similarly the space gap dynamics can be written in discrete time as:
\begin{equation}
\begin{array}{ll}
\label{eq:spacing dynamic}
      s_{k+1} = & s_{k} + \Delta T (v_{l,k}-v_{k}).
      \end{array}
\end{equation}

Grouping together~\eqref{eq:CTH-FTL-Discrete} and~\eqref{eq:spacing dynamic}, the model can be rewritten as a state space model of the form:
\begin{equation}\label{eq:DiscreteMappingFormula}
    \boldsymbol{x}_{k+1} = \boldsymbol{A} \boldsymbol{x}_{k} + \boldsymbol{B} \boldsymbol{u}_{k}
\end{equation}
\noindent where $\boldsymbol{x}_{k} \in \mathbb{R}^{2}$ is the state vector at timestep ${k}$, with $\boldsymbol{x}_{k} = [v_{k}, s_{k}]^{T}$. Additionally, $\boldsymbol{u}_{k} \in \mathbb{R}^{1}$ is an input to the system, which in this case is the lead vehicle speed $\boldsymbol{u}_{k} = [v_{l,k}]$. The matrix $\boldsymbol{A} \in \mathbb{R}^{2\times2}$ is the dynamics matrix and describes how the state evolves according to $v_{k}$ and $s_{k}$, and $\boldsymbol{B} \in \mathbb{R}^{2\times1}$ is the control dynamics vector which accounts for the impact of the input, i.e., from $v_{l,k}$.

A straightforward arrangement of~\eqref{eq:CTH-FTL-Discrete} and~\eqref{eq:spacing dynamic} into \eqref{eq:DiscreteMappingFormula} gives $\boldsymbol{A}$ and $\boldsymbol{B}$ in terms of the car following parameters as:
\begin{equation}\label{eq:State_Space_Values}
\boldsymbol{A} = 
\begin{bmatrix}
    1-(k_1\tau +k_2)\Delta T &  k_{1} \Delta T \\
    -\Delta T & 1 \\
\end{bmatrix}, 
\quad
\boldsymbol{B} = 
\begin{bmatrix}
    k_{2}\Delta T\\
    \Delta T\\
\end{bmatrix}.
\end{equation}
For forward simulation, given an initial condition and the parameters in~\eqref{eq:State_Space_Values}, the model ~\eqref{eq:DiscreteMappingFormula} can be used to describe the trajectory evolution of the ACC vehicle.

Since the motivation of this work is to propose methods to recover car-following model parameters from experimental data, we need to solve an inverse problem. Concretely, we are given a timeseries of direct measurements of the state $\boldsymbol{x}_k$ and input $\boldsymbol{u}_k$, and we wish to recover the parameters in \eqref{eq:CTH-FL}. To simplify the problem we recognize that if we directly estimate the $i,j$ entries of the matrices $\boldsymbol{A}$ and $\boldsymbol{B}$ (denoted $a_{i,j}$ and $b_{i,j}$), we can compute the car following model parameters as:  
\begin{equation}
\label{eq:var_change}
\begin{array}{cc}
     & k_{1} = \frac{a_{1,2}}{\Delta T} \quad k_2 = \frac{b_{1,1}}{\Delta T} \quad \tau = \frac{1-b_{1,1}-a_{1,1}}{a_{1,2}}\\
\end{array}
\end{equation}

We now show how to estimate the entries of the matrices $\boldsymbol{A}$ and $\boldsymbol{B}$ via least-squares. First, for a dataset with $N$ samples each of speed, space gap, and relative velocity collected at a uniform sample rate, define:
$$
\boldsymbol{X} = 
\begin{bmatrix}
    v_{1}& \dots & v_{N-1} \\
    s_{1}& \dots & s_{N-1} \\
\end{bmatrix},
\quad
\boldsymbol{U} = 
\begin{bmatrix}
    v_{l,1}& \dots & v_{l,N-1} \\
\end{bmatrix}, 
\quad
\boldsymbol{X'} = 
\begin{bmatrix}
    v_{2}& \dots & v_{N} \\
    s_{2}& \dots & s_{N} \\
\end{bmatrix},
$$

\noindent such that $\boldsymbol{X}$ contains measurements of $v_{k}$ and $s_{k}$ from timestep 1 to $N-1$ in column-wise order. The matrix $\boldsymbol{U}$ contains all $v_{l,k}$ from 1 to $N-1$, which are the control inputs to \eqref{eq:DiscreteMappingFormula}. The term $\boldsymbol{X'}$ contains the values of $v_{k}$ and $s_{k}$ from timestep 2 to $N$. Given the data matrices $\boldsymbol{X}$, $\boldsymbol{X'}$, and $\boldsymbol{U}$, \eqref{eq:DiscreteMappingFormula} can now be rewritten as the following system of equations:

\begin{equation}
\label{eq:LS_no_error}
\boldsymbol{X'} =
    \boldsymbol{A}\boldsymbol{X} + \boldsymbol{B}\boldsymbol{U},\\
\end{equation}
in which the unknowns are the entries of $\boldsymbol{A}$ (technically only containing 2 unknowns since $a_{2,2}=1$ and $a_{2,1}=-\Delta T$ ), and $\boldsymbol{B}$ (note also, $b_{1,2}=\Delta T$).
In the case when the system has many more equations than unknowns (i.e., as is the case when any nontrivial dataset is collected), and when the measurements contain errors, then~\eqref{eq:LS_no_error} holds only approximately:
$$
    \boldsymbol{X'} \approx
    \boldsymbol{A}\boldsymbol{X} + \boldsymbol{B}\boldsymbol{U}\\
$$

 In order to solve for the parameter set that minimizes the error (in the sense of the sum of squared errors), the optimal matrices $\boldsymbol{A^{*}}$ and $\boldsymbol{B^{*}}$ can be found by solving the following least-squares problem:
\begin{equation}\label{eq:LeastSquaresFormula_Frobenius}
    \boldsymbol{A^{*}},\boldsymbol{B^{*}} = \underset{\boldsymbol{A},\boldsymbol{B}}{{\text{argmin}}}: \left\Vert\ \boldsymbol{X'} - \boldsymbol{A} \boldsymbol{X} - \boldsymbol{B} \boldsymbol{U}\rm\ \right\Vert_{F}\\
\end{equation}
\noindent where $\boldsymbol{A^{*}}$ and $\boldsymbol{B^{*}}$ minimize the Frobenius Norm (denoted $\left\Vert \cdot \right\Vert_{F}$) of the error between the proposed dynamics and the real data. This is equivalent to solving the problem:
\begin{equation}\label{eq:eq:LeastSquaresFormula_Sum}
    \boldsymbol{A^{*}},\boldsymbol{B^{*}} =\underset{\boldsymbol{A},\boldsymbol{B}}{{\text{argmin}}}: \sqrt{\sum_{k=1}^{N-1} \left\Vert\left( \boldsymbol{x_{k+1}} - \boldsymbol{A} \boldsymbol{x_{k}} - \boldsymbol{B} \boldsymbol{u_{k}}\rm\right) \right\Vert^{2}_{2}}.\\
\end{equation}

\noindent Finally problem~\eqref{eq:eq:LeastSquaresFormula_Sum} (equivalently~\eqref{eq:LeastSquaresFormula_Frobenius}) can be solved explicitly via a pseudo-inverse:

$$
\begin{bmatrix}
    \boldsymbol{A^{*}} & \boldsymbol{B^{*}} \\
\end{bmatrix}
= 
\begin{bmatrix}
\boldsymbol{X'} 
\end{bmatrix}
\begin{bmatrix}
    \boldsymbol{X} \\
    \boldsymbol{U} \\
\end{bmatrix}^{\dagger},\\
$$
and the resulting model parameters are recovered via~\eqref{eq:var_change}.

\subsection{Online Estimation Problem}\label{sec:Particle_Filter_Seciont}
The parameter estimation problem can also be framed as an online estimation problem, in which a model is learned in a single sweep through the data. Such methods, if fast enough, may be used for real-time processing of data in order to estimate the model parameters during data collection. Real-time execution of an online model can fix issues with offline methods which may have increasing complexity with large problems, as they execute as data is collected. Additionally, having an active model estimate of a driver's behavior may allow for control objectives that take advantage of the knowledge of this model. As such, accurate online model estimation methods are of general interest. 

A sequential parameter estimation approach using the particle filter is outlined in this section. Because the filtering problem treats both the typical state variables and the parameters as part of the augmented state to be estimated, a nonlinear estimator such as the particle filter is needed.

In order to learn the model parameters, $\boldsymbol{\theta}$ = $[k_1,k_2,\tau]^T$, a joint parameter and state estimation problem is posed. First, $\boldsymbol{\theta}$ is concatenated to the system state $\boldsymbol{x}_k \in \mathbb{R}^2$ to form an augmented state $\mathcal{X}_k \in \mathbb{R}^5$. The augmented state vector is then sequentially estimated from the measurements, producing estimates of the parameters. 

The system in state space from is written as:
\begin{equation}\label{state_space_model}
\begin{gathered}
    \mathcal{X}_k = 
    \begin{bmatrix}
        \boldsymbol{x}_k \\
        \boldsymbol{\theta}_k
    \end{bmatrix} = 
    \begin{bmatrix}
        f(\boldsymbol{\theta}_{k-1}, \boldsymbol{x}_{k-1}, \boldsymbol{u}_{k-1})\\
        \boldsymbol{\theta}_{k-1}
    \end{bmatrix} +
    \boldsymbol{w}_{k-1}   \in \mathbb{R}^5\\
    \boldsymbol{y}_{k} = h(\boldsymbol{x}_{k}) + \boldsymbol{v}_{k}  \in \mathbb{R}^2,
\end{gathered}
\end{equation}
where $f$ (with a slight abuse of notation) represents the discretized car-following model described in~\eqref{eq:CTH-FL}; $h(\cdot)$ is the measurement equation, which in this case maps the augmented state vector to the measurements taken by the on-board radar sensor; $\boldsymbol{w}_{k} \sim(0,Q) \in \mathbb{R}^5$ and $\boldsymbol{v}_{k} \sim(0,R) \in \mathbb{R}^2$ are the additive process noise and measurement noise distributions at time $k$, respectively, and $Q \in \mathbb{R}^{5\times5}$ and $R \in \mathbb{R}^{2\times2}$ are the process and measurement error covariance matrices. The particle filter, among other filtering techniques, is deployed because of its flexibility in noise distribution and its relaxed assumption about the linearity of the dynamics of the system. More details on the particle filter implementation can be found in standard references such as~\cite{simon}.

The particle filter uses weighted particles (samples) approximate the conditional state distribution given all measurements up to the current timestep using a sequential estimation approach. Therefore, the output is a probability distribution for each parameter at each time step instead of a single point estimate, in contrast to the other two methods presented in this work.

\section{Estimation On Synthetic Data}\label{sec:SynthCal}
In this section, each parameter estimation routine described above is run on synthetically generated data. This is done to understand the ability for each to recover the true model parameters with no sensor noise. The data is created by selecting a set of model parameters and a predefined lead driver speed profiles. A time-series of speed and spacing values using \eqref{eq:CTH-FTL-Discrete} is then created via a forward Euler time-stepping scheme at the same frequency that the data is collected. The simulated data is then fed into each estimation method, with each returning a set of estimated parameter values. It is found that the two offline models are able to exactly recover the model parameters, while the online estimation finds a model close to it. Additionally, the run-times for each method are calculated and discussed.

\subsection{Synthetic Data Creation}
In order to create a set of synthetic data, the car-following model~\eqref{eq:CTH-FL} is simulated given a known set of parameters to be recovered by the estimator. In the experiment below, the true parameters used are: $\boldsymbol{\theta}_{true} = [0.08, 0.12, 1.5]^T$. These values are representative of parameters that have been reported in ~\cite{milanes2014modeling,gunter2019are}. Additionally, in order to generate a synthetic dataset, the speed profile of a lead vehicle is needed, along with an initial spacing, and the initial following vehicle's speed. These values were all taken from~\cite{gunter2019are}. In this data a driver engaged in rapid deceleration and subsequent acceleration events, interspersed with steady driving periods for a span of 620 seconds. The lead vehicle's driving profile was chosen because it is known to reflect a physically possible driving profile, and contains both steady-state driving and dynamic behavior.

\begin{figure}
    \centering
    \includegraphics[width=\columnwidth]{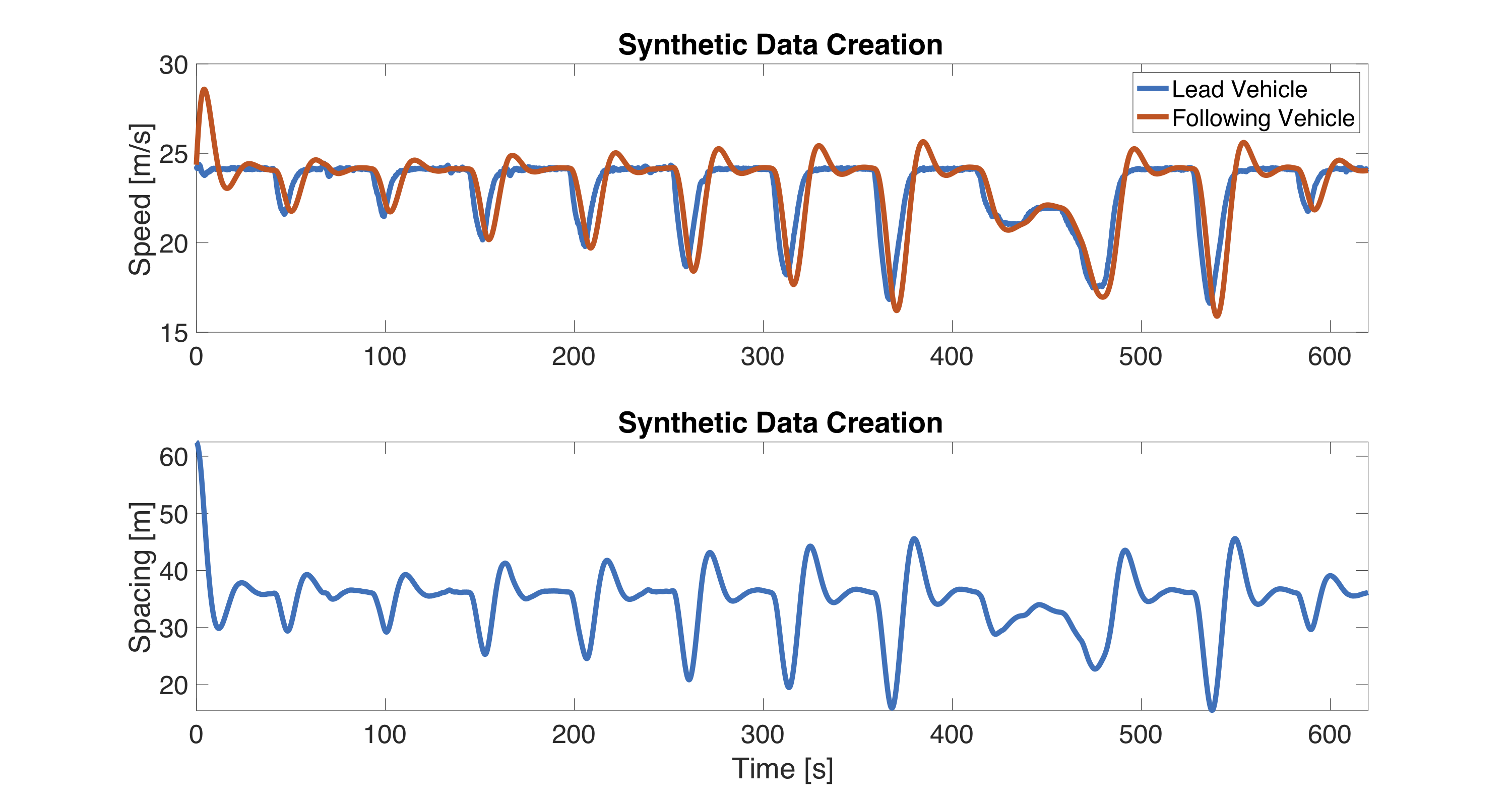}
    \caption{Synthetic Data used for parameter estimation. }
    \label{fig:SynthData}
\end{figure}

In order to perform the simulation the time-stepping routine outlined in \eqref{eq:DiscreteMappingFormula} is used to create time-series data over the same length of time as the lead vehicle trajectory. First, the simulation is initialized at a follower speed and spacing of $24.4$ m/s and $62.5$ m, which are also taken from ~\cite{gunter2019are}. The data is sampled at a rate of 10 hertz, meaning that the timestep is $\Delta T = 0.1$ seconds. From these initial conditions the speed and inter-vehicle space gap are then determined via simulation of~\eqref{eq:CTH-FTL-Discrete}. Figure~\ref{fig:SynthData} displays the lead vehicle and follower speed profiles and the space gap data. 

The true parameters used to generate the synthetic dataset corresponds to a string-unstable model, meaning that it will tend to amplify lead vehicle disturbances. This can be seen in the simulation data, as in several occasions the leader vehicle slows down more than the leader, and speeds up more than the leader. 

\subsection{Estimation Results}\label{sec:SynthCal_Results}
Using the synthetic data created above, we now turn to the results of each parameter estimation routine that attempt to recover the true parameters using only the measurement data. We use the \textit{mean absolute error} (MAE) in spacing and speed to compare the accuracy of each estimation method. The particle filter requires several algorithm parameters to be set, which are summarized in Table~ \ref{tab:pf_param}. This includes the number of particles used in the estimator $N_p$, the initial distribution of the augmented state vector, which is assumed to follow a normal distribution with mean $\boldsymbol{\mu}_{x_0}$ and covariance $Q_0$, and the model and measurement covariances $Q$ and $R$.

\begin{table}
\centering
\begin{tabular}{@{}ll@{}}
\toprule
\text{Parameters} & \text{Values} \\ \midrule
$N_p$ & 500 particles \\
$\boldsymbol{\mu}_{x_0}$ & [s(0), v(0), 0.1, 0.1, 1.4] \\
$Q_0$ & diag{[}0.5m 0.5m/s $0.2s^{-2}$ $0.2s^{-1}$ 0.3s{]} \\
$Q$ & diag{[}0.2m  0.1m/s  $0.01s^{-1}$  $0.01s^{-1}$  0.01s{]} \\
$R$ & diag{[}0.2m  0.1m/s{]} \\ \bottomrule
\end{tabular}
\caption{Particle filter parameters}
\label{tab:pf_param}
\end{table}

Results of the estimation are shown in Table~\ref{tab:synth_performance_summary}, displaying the runtime, MAE, and finally the corresponding string-stability estimate based on the parameters recovered by the estimator. The batch method recovers the exact model parameters used. Overall the method takes 19.3 seconds to complete. The least-squares approach also recovers exactly the parameters used to generate the synthetic data, while completing completely in 0.0094 seconds. This is the best runtime of the three models by about 5 orders of magnitude. The particle filter takes the longest of the three methods at 49 seconds. While the particle filter takes the longest of the three methods, it should be noted that since it finishes its computation well under the length of the experiment, it may feasibly be executed as a true real-time estimation method.

\begin{table}
\centering
\begin{tabular}{@{}cccc@{}}
\toprule
\text{Criteria}                                              & \text{Batch optimization}                                             & \text{Least squares}                                                   & \text{Particle filter}                                                \\ \midrule
\begin{tabular}[c]{@{}c@{}}Estimated \\ parameter \\ values\end{tabular} & \begin{tabular}[c]{@{}c@{}}$k_1 = 0.08$\\ $k_2 = 0.12$\\ $\tau = 1.5$ \end{tabular} & \begin{tabular}[c]{@{}c@{}}$k_1 = 0.08$\\ $k_2 = 0.12$\\ $\tau = 1.5$
\end{tabular} & \begin{tabular}[c]{@{}c@{}}$k_1 = 0.041$\\ $k_2 = 0.21$\\ $\tau = 1.41$\end{tabular} \\ \midrule
Algorithm                                                      & Offline                                                                 & Offline                                                                 & Online                                                                  \\ \midrule
Running time (s)                                               &                    19.30                                                     &                        0.00940                                                &    49                                                                     \\ \midrule
MAE Spacing (m)                                               &     0                                                                   &         0                                                                &     2.544                                                                    \\ \midrule
MAE Speed ($\frac{m}{s}$)                                               &        0                                                                 &                 0                                                        &     0.3184                                                                    \\ \midrule
String stability                                               &  string unstable                                                                       &  string unstable                                                                       & 98.52\% likely to be string unstable                                           \\ \bottomrule
\end{tabular}
\caption{Performance summary of all estimation methods on synthetic data.}
\label{tab:synth_performance_summary}
\end{table}

The particle filter recovers a distribution on the model parameters with a mean value that is close to the true parameters. When simulating with the mean values of the parameter estimates, the calibrated model has an MAE of $0.3184$ m/s in speed and $2.544$ m in the spacing. Given that the particle filter assumes model and sensor noise (see $Q$ and $R$ from the particle filter section) it is not surprising that the mean value of the parameter estimate does not perfectly recover the true parameters (since the model and measurement in fact have zero error in this synthetic example). 

It is additionally worth noting that that particle filter provides a distribution for the model parameters by looking at what values are present after the resampling stage of the particle filter algorithm. In this case, a final parameter set is chosen by selecting the mean of the particle distribution. This distribution can give a probability of the model being string unstable and gives information about the range of parameters. This is done by checking the string stability of each particle after the re-sampling step and calculating the percentage which have string unstable parameters. In the above example it is found that $98.52\%$ of the particles yield a string unstable model.

\section{Estimation On Real Data}
\label{sec:RealCal}
In this section each estimation method is run on real data collected from an ACC equipped vehicle. Experimental details are outlined for how data is collected for estimation, and the results of the estimation are presented. First the experimental setup is outlined, describing how speed and spacing data are collected from an ACC vehicle's radar system. Next the data is compared with GPS data that was simultaneously collected during the experiment. A \textit{U-blox evaluation kit} GPS platform with sub-meter accuracy is used as a benchmark to determine the accuracy of the ACC vehicle's radar data. It is observed that the radar measurements closely match the GPS devices, indicating that the radar data is a viable alternative to the GPS data. Additionally, this means that ACC radar may be used in general as a method to collect vehicle trajectory data, without the need to instrument two vehicles to collect spacing data from GPS.

Next the results of the different estimation routines are displayed. Each estimation method produces a different set of parameters when calibrated on the observed data. MAE values in both speed and spacing are used to examine the accuracy of each method to reproduce observed data, and all models are found to have similar MAE values that are considered low and are similar to those found in the works~\cite{gunter2019arxiv,gunter2019are,milanes2014modeling}. Additionally, the distributions of speed and spacing error are plotted as histograms. The speed errors for each calibrated model are very close to zero-centered, and while the spacing errors for the least-squares and particle filter methods are not zero-centered they are not significantly far from it. The conclusion is that due to both the low MAE values associated with each model and their error distributions that each estimation method is able to converge to a model of good accuracy to the recorded driving behavior. 
\subsection{Experimental Details}\label{sec:ExpMethodology}
Inter-vehicle spacing and relative speed difference data is gathered from a lead and following vehicle. The following vehicle is an ACC equipped vehicle whose parameters are to be estimated directly from the radar data. The vehicle used in this experiment is a commercially available 2019 SUV with a full speed range adaptive cruise control system. A total of 900 seconds (15 minutes) of data are continuously recorded in which the ACC vehicle drives in traffic on a freeway in Nashville, TN following a leader equipped with GPS (used for benchmarking the radar sensor accuracy). The lead driver simply drives according to their normal driving behavior, while the ACC vehicle follows in ACC mode. The entire experiment was completed with continuous following from the ACC vehicle, as in no de-activations in the ACC were required and no cars cut in between drivers. 

Data is collected by recording measurements from the CAN bus on the vehicle, which contains speed, space gap, and relative velocity data. The radar unit reports both inter-vehicle spacing between the two vehicles and speed differences between the two over the course of the experiment. The speedometer reports ACC vehicle's speed on the CAN bus.

\begin{figure}
    \centering
    \includegraphics[width=\columnwidth]{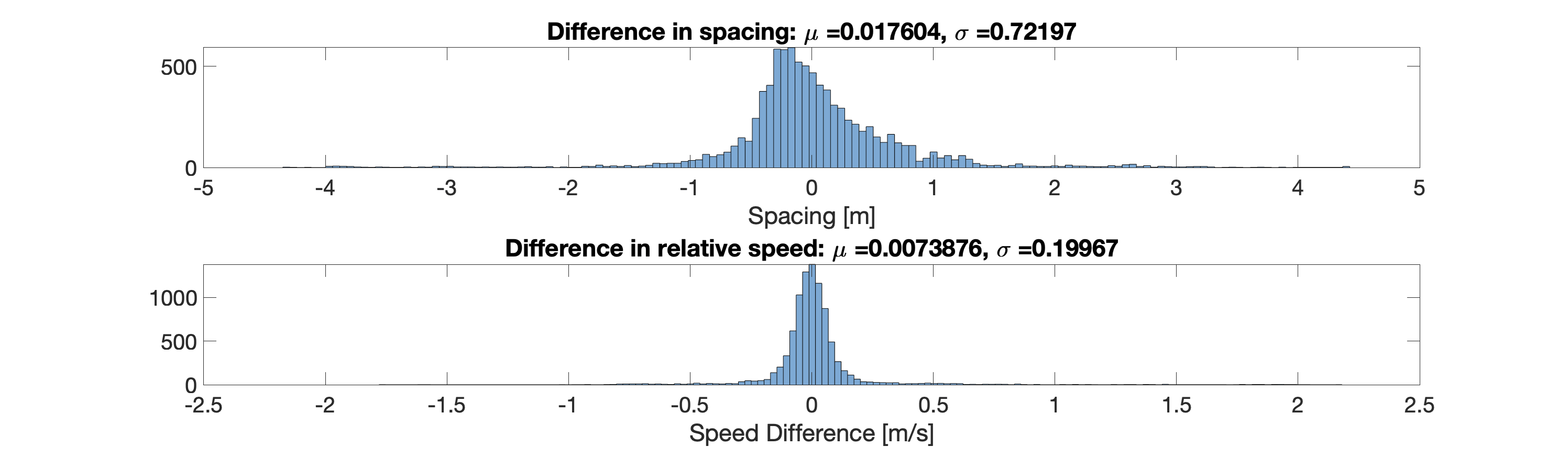}
    \caption{Histogram of the difference between GPS measureuments and CAN bus measurements for the inter-vehicle spacing and the relative speed difference.}
    \label{fig:Histogram}
\end{figure}

In order to verify the accuracy of the vehicle's radar unit both vehicles are additionally equipped with sub-meter accurate GPS units which track global position and speed. As a result, two timeseries of spacing and speed are recorded and compared for similarities. A histogram of error between the two measurement techniques is displayed in figure \ref{fig:Histogram}.
{}

Both the difference between spacing measurements and relative speed measurements for the radar compared to the GPS are close to zero-centered. This suggests that there is general agreement between the two different sensors. Additionally, the standard deviations of the two errors ($0.70$ m and $0.20$ m/s for spacing and speed respectively) are considered to be quite low, suggesting that neither suffers from much sensor noise. In order to further verify that the radar device produces reasonable values for the desired purpose each estimation routine was also run on the respective GPS data. In each case the methods converged to models very similar to those displayed in the next section. From this it is deemed that the on-board radar measurements from the vehicle itself are of suitable quality for use in this study, and in general for measuring the data necessary to calibrate a car-following model.

\subsection{Estimation Results}
To compare the performance of each method outlined above, the dataset outlined previously that contains speed, spacing, and speed difference estimates is fed into the different parameter estimation methods. In order to compare the performances of each model the MAE between the simulated speed and real speed, and simulated spacing and real spacing are calculated for each model. Additionally, the run-time for each model to complete is calculated. The string stability estimate that each method returns is also shown. These results are displayed in Table \ref{tab:real_performance_summary}.

\begin{table}
\centering
\begin{tabular}{@{}cccc@{}}
\toprule
\text{Criteria}                                              & \text{Batch optimization}                                             & \text{Least-squares}                                                   & \text{Particle filter}                                                \\ \midrule
\begin{tabular}[c]{@{}c@{}}Estimated \\ parameter \\ values\end{tabular} & \begin{tabular}[c]{@{}c@{}}$k_1 = 0.0227$\\ $k_2 = 0.194$\\ $\tau = 1.227$ \end{tabular} & \begin{tabular}[c]{@{}c@{}}$k_1 = 0.0174 $\\ $k_2 = 0.1641$\\ $\tau = 1.127$
\end{tabular} & \begin{tabular}[c]{@{}c@{}}$k_1 = 0.0353$\\ $k_2 = 0.2241$\\ $\tau = 1.2608$\end{tabular} \\ \midrule
Algorithm                                                      & Offline                                                                 & Offline                                                                 & Online                                                                  \\ \midrule
Running time (s)                                               &          18.59                                                               &       0.055                                                                  &    73.5                                                                     \\ \midrule
MAE Speed (m)                                               &          0.2384                                                               &           0.2626                                                              &     0.2916                                                                  \\ \midrule
MAE Spacing ($\frac{m}{s}$)                                               &        2.0243                                                                 &          3.5556                            &           2.4478                                                            \\ \midrule
String stability                                               &  string unstable                                                                       &  string unstable                                                                       & 88.65\% likely to be string unstable                                           \\ \bottomrule
\end{tabular}
\caption{Performance summary of all estimation methods on ACC data.}
\label{tab:real_performance_summary}
\end{table}
The least-squares estimation routine again produces the fastest computation time at 0.055 seconds, with the batch optimization achieving 18.6 seconds and the particle filter executing in 73.5 seconds. The least-squares estimation approach executes roughly 3 orders of magnitude faster than the other two methods. While length of the simulation increased by 47$\%$ compared to the data presented in \ref{sec:SynthCal_Results}, the batch optimization scheme executed 4$\%$. The particle-filter in comparison executed 50$\%$ faster than in the previous example. Comparatively, the least-squares method increased by 485$\%$, which can be attributed to the quadratic runtime complexity of the pseudo-inverse calculation. The particle filter again runs in less time than the total time of the experiment and is an online method, meaning it could be run as a real-time estimation algorithm. 

\begin{figure}
    \centering
    \includegraphics[width=\columnwidth]{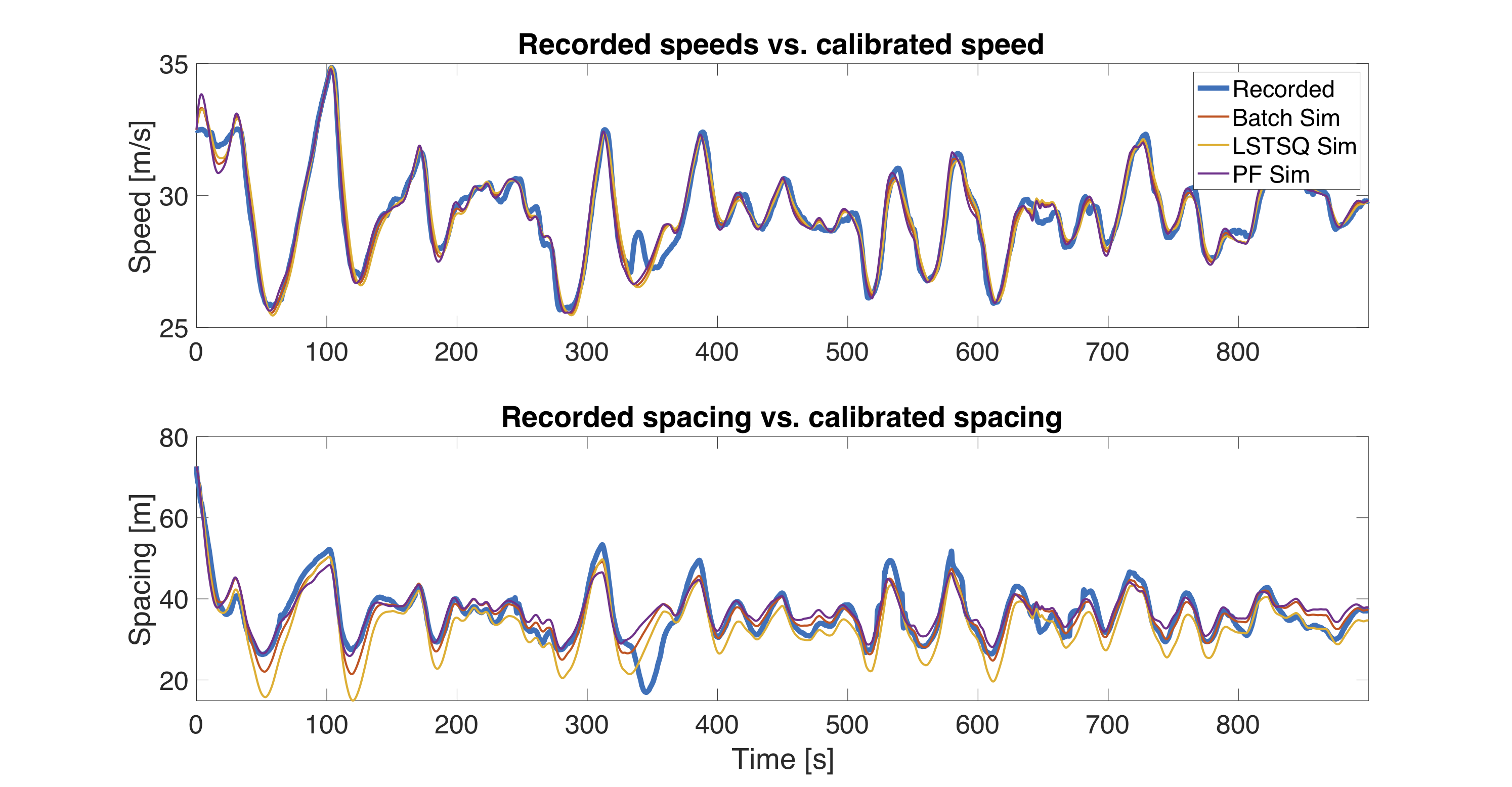}
    \caption{Comparison between recorded vehicle speed and spacing vs simulated for each model found.}
    \label{fig:ModelComp}
\end{figure}

The batch optimization converges to the model that achieves both the lowest MAE speed and spacing errors at $0.2384$ m/s and $2.0243$ m. This represents percent errors of $0.81 \%$ in speed and $5.6 \%$ in spacing. The least-squares method converges to a model with MAE values of $0.2626$ m/s and $3.5556$ m, which are percent errors of $0.9 \%$ in speed and $10.7 \%$ in spacing. Finally, the particle filter routine finds a model with MAE values of $0.2916$ m/s and $2.4478$ m, which correspond to percent errors of $0.90 \%$ in speed and $6.7 \%$ in spacing. In general each method converges to a calibrated model with similar MAEs in speed and spacing, and similar percent errors. Additionally, these models are similar in scale to those found in other works ~\cite{milanes2014modeling,gunter2019arxiv,gunter2019are}. These results suggest that each method was able to recover a model that reproduces good prediction of the ACC driving behavior. A time-series simulation for each model is presented in Figure \ref{fig:ModelComp}.

\begin{figure}
    \centering
    \includegraphics[width=\columnwidth]{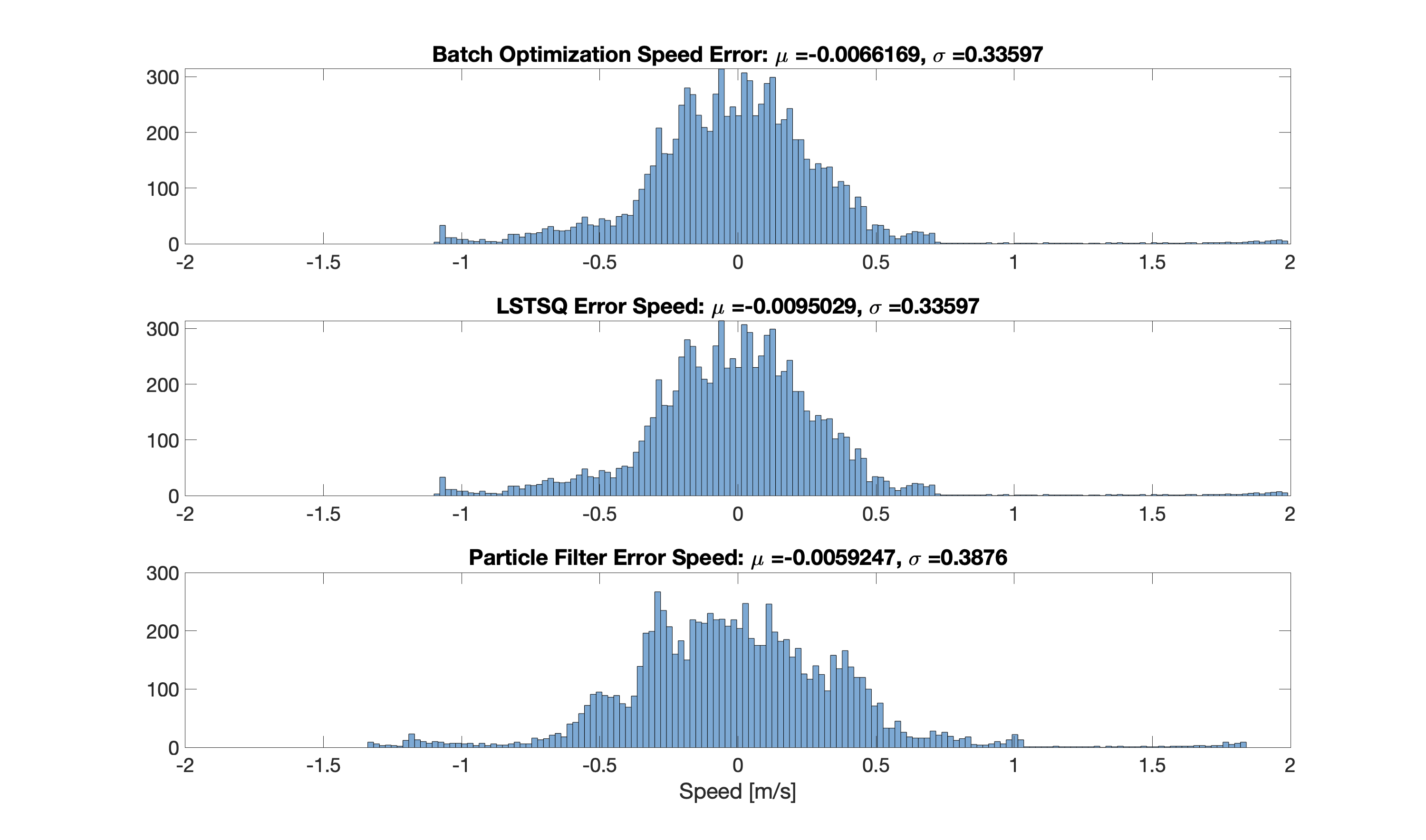}
    \caption{Error Distributions in speed for each calibrated model.}
    \label{fig:Speed_Hist_Comp}
\end{figure}

\begin{figure}
    \centering
    \includegraphics[width=\columnwidth]{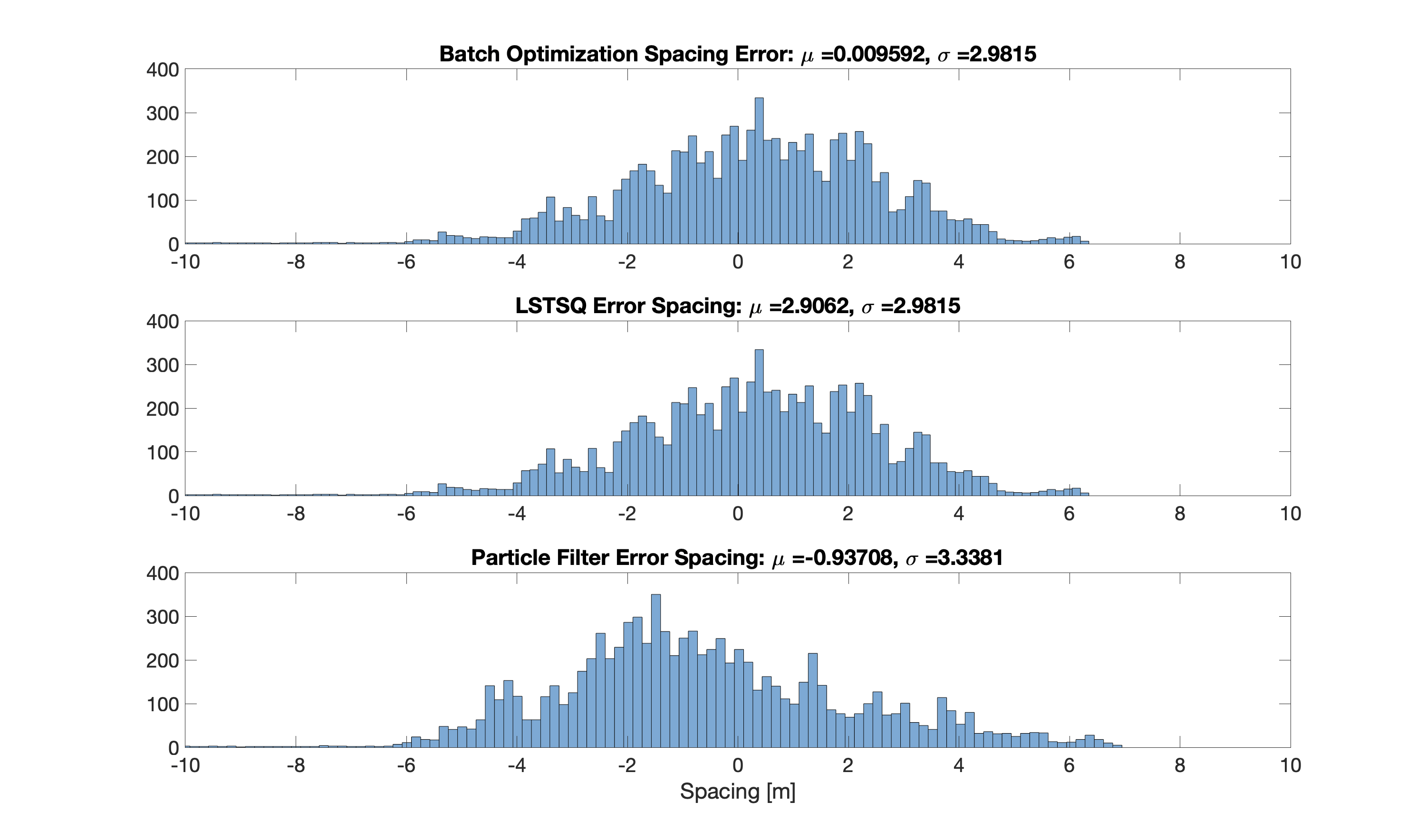}
    \caption{Error Distributions in spacing for each calibrated model.}
    \label{fig:Spacing_Hist_Comp}
\end{figure}

In general one can note the goodness of fit of the three models. Each calibrated model produces a similar speed profile, as is to be expected given the low MAE values on speed for each model. A small exception is during the period between roughly 325 seconds and and 375 seconds, in which the vehicle engages in an acceleration which none of the calibrated models is able to capture. This underscores that while each model produces a good recreation of the observed behavior, they cannot perfectly describe the potentially non-linear dynamics of the vehicle's control and dynamics. This underlies that it would not be realistic to expect that the proposed method could under any estimation routine achieve completely correct modeling of the ACC system, and as such errors should be expected. To further understand the goodness of fit of these models the histograms of the errors are shown in \ref{fig:Speed_Hist_Comp} and \ref{fig:Spacing_Hist_Comp}.

For all three models the average error is less than $0.01$ m/s with standard deviations under $0.4$ m/s. The batch optimization Scheme returns a model that has an average spacing error of within $0.01$ m, while least-squares returns a model that has an average spacing of $2.9$ m less than the real ACC, and the particle filter returns a model that has on average a model that is $0.94$ m more than the ACC. Each model has an spacing error distribution with a standard deviation less than $3.4 m$, but more than $2.9$ m. Given that the least-squares approach and the particle-filter have average spacing errors which are significantly different from zero, this suggests that there is some bias in the spacing values the model predicts, but given than all models have average speed errors which are very close to zero there does not appear to be any bias in the speed estimates. While there may be some bias in the spacing estimates of the models from the least-squares and the particle-filter the magnitude of the bias is small. 

Given the results presented in this section, it can be concluded that each model achieves small MAE values compared to the recorded data. These MAE values and the corresponding percentage errors are also of the same order as is reported in other works examining this problem ~\cite{milanes2014modeling,gunter2019are,gunter2019arxiv}. Additionally, all models have either effectively zero bias or negligible biases in the error, also suggesting the models are good recreations of the ACC driving behavior.

\section{Conclusion}\label{sec:Conclusion}
 Three car-following estimation routines are outlined and their performance on calibrating both with respect to synthetically calculated data and real driving data is assessed. Models are calibrated using data recorded directly from and ACC vehicle's on-board radar system. Both the batch and linear least-squares are able to recover an underlying model given no sensor or model noise, while the particle filter method comes close but does not converge exactly. All methods were found to return model parameters which correspond to low MAE values in simulation for both speed and spacing. Additionally the models were found to have error distributions with minimal bias, also suggesting goodness of fit. For these reasons the models are considered to be good recreations of the ACC driving behavior, and the three estimation methods are all considered to work well at identifying the ACC system. 

In general the least-squares method is well suited to rapid parameter estimation of linear car following models for modest datasets. The batch optimization method has as good an ability to fit an accurate model as the least-squares method, but can scale to non-linear car-following models as well. Finally, the particle filter was the slowest method but recovers similar parameter values to the two offline methods. It is also able to complete calculations faster than the length of of the dataset used, meaning it could be employed as a real-time online estimation technique. 

In future work, the online estimation methods will be expanded by developing the least-squares estimation problem into a real-time estimation method. Additionally, further techniques for car-following model calibration can be considered which leverage the fact that ACC vehicles report an estimated acceleration over the CAN-bus, in conjunction with speed and space-gap values.

\section{Acknowledgements}
This material is based upon work supported by the National Science Foundation under Grant No. CMMI-1853913 (Wang) and CNS-1837652 (Gunter).

\section{Author contributions}
Y. Wang developed and wrote material related to the particle filter. G. Gunter developed and wrote material related to the least-squares. M. Nice developed material related to the data collected from the vehicle's CAN bus. Y. Wang and G. Gunter contributed equally to the writing of manuscript and experimental collection. D. Work supervised the research and assisted with manuscript preparation.
\bibliographystyle{unsrt}
\bibliography{main_arxiv}

\begin{thebibliography}{10}

\bibitem{gunter2019are}
G.~Gunter, D.~Gloudemans, R.~E. Stern, S.~McQuade, R.~Bhadani, M.~Bunting,
  M.~L. Delle~Monache, R.~Lysecky, B.~Seibold, J.~Sprinkle, B.~Piccoli, and
  D.~B. Work.
\newblock Are commercially implemented adaptive cruise control systems string
  stable?
\newblock {\em arXiv preprint arXiv:1905.02108}, 2019.

\bibitem{gunter2019arxiv}
G.~Gunter, C.~Janssen, W.~Barbour, R.~E. Stern, and D.~B. Work.
\newblock Model based string stability of adaptive cruise control systems using
  field data.
\newblock {\em arXiv preprint arXiv:1902.04983}, 2019.

\bibitem{milanes2014modeling}
V.~Milan{\'e}s and S.~E. Shladover.
\newblock Modeling cooperative and autonomous adaptive cruise control dynamic
  responses using experimental data.
\newblock {\em Transportation Research Part C: Emerging Technologies},
  48:285--300, 2014.

\bibitem{knoop2019platoon}
V.~L. Knoop, M.~Wang, I.~Wilmink, D.~M. Hoedemaeker, M.~Maaskant, and
  EJ~Van~der Meer.
\newblock Platoon of sae level-2 automated vehicles on public roads: Setup,
  traffic interactions, and stability.
\newblock {\em Transportation Research Record}, page 0361198119845885, 2019.

\bibitem{kesting2010enhanced}
Arne Kesting, Martin Treiber, and Dirk Helbing.
\newblock Enhanced intelligent driver model to access the impact of driving
  strategies on traffic capacity.
\newblock {\em Philosophical Transactions of the Royal Society of London A:
  Mathematical, Physical and Engineering Sciences}, 368(1928):4585--4605, 2010.

\bibitem{gipps1981behavioural}
P.~G. Gipps.
\newblock A behavioural car-following model for computer simulation.
\newblock {\em Transportation Research Part B: Methodological}, 15(2):105--111,
  1981.

\bibitem{BandoHesebeNakayama1995}
M.~Bando, K.~Hesebem, A.~Nakayama, A.~Shibata, and Y.~Sugiyama.
\newblock Dynamical model of traffic congestion and numerical simulation.
\newblock {\em Physical Review E}, 51(2):1035--1042, 1995.

\bibitem{MobileCentury}
J.~C. Herrera, D.~B. Work, R.~Herring, X.~Ban, Q.~Jacobson, and A.~M. Bayen.
\newblock Evaluation of traffic data obtained via gps-enabled mobile phones:
  The mobile century field experiment.
\newblock {\em Transportation Research Part C: Emerging Technologies},
  18(4):568 -- 583, 2010.

\bibitem{NGSimCalibration1}
C.~{Chen}, L.~{Li}, J.~{Hu}, and C.~{Geng}.
\newblock Calibration of mitsim and idm car-following model based on ngsim
  trajectory datasets.
\newblock In {\em Proceedings of 2010 IEEE International Conference on
  Vehicular Electronics and Safety}, pages 48--53, July 2010.

\bibitem{NGSimCalibration2}
A.~Duret, C.~Buisson, and N.~Chiabaut.
\newblock Estimating individual speed-spacing relationship and assessing
  ability of newell's car-following model to reproduce trajectories.
\newblock {\em Transportation Research Record}, 2088(1):188--197, 2008.

\bibitem{NGSimCalibration3}
V.~{Punzo}, M.~{Montanino}, and B.~{Ciuffo}.
\newblock Do we really need to calibrate all the parameters? variance-based
  sensitivity analysis to simplify microscopic traffic flow models.
\newblock {\em IEEE Transactions on Intelligent Transportation Systems},
  16(1):184--193, Feb 2015.

\bibitem{KestingModelCalibration2008}
A.~Kesting and M.~Treiber.
\newblock Calibrating car-following models by using trajectory data:
  Methodological study.
\newblock {\em Transportation Research Record}, 2088(1):148--156, 2008.

\bibitem{PunzoModelCalibration2005}
V.~Punzo and F.~Simonelli.
\newblock Analysis and comparison of microscopic traffic flow models with real
  traffic microscopic data.
\newblock {\em Transportation Research Record}, 1934(1):53--63, 2005.

\bibitem{Monteil2015RealTimeEO}
J.~Monteil, N.~O'Hara, V.~Cahill, and M.~Bouroche.
\newblock Real-time estimation of drivers' behaviour.
\newblock {\em 2015 IEEE 18th International Conference on Intelligent
  Transportation Systems}, pages 2046--2052, 2015.

\bibitem{Monteil_identifiability}
J.~{Monteil} and M.~{Bouroche}.
\newblock Robust parameter estimation of car-following models considering
  practical non-identifiability.
\newblock In {\em 2016 IEEE 19th International Conference on Intelligent
  Transportation Systems (ITSC)}, pages 581--588, Nov 2016.

\bibitem{Beckmann_tuning}
D.~{Beckmann}, M.~H. {Riva}, M.~{Dagen}, and T.~{Ortmaier}.
\newblock Comparison of online-parameter estimation methods applied to a linear
  belt drive system.
\newblock In {\em 2016 European Control Conference (ECC)}, pages 364--369, June
  2016.

\bibitem{AutomatedTraffic_1}
P.~{Ioannou}, Z.~{Xu}, S.~{Eckert}, D.~{Clemons}, and T.~{Sieja}.
\newblock Intelligent cruise control: theory and experiment.
\newblock In {\em Proceedings of 32nd IEEE Conference on Decision and Control},
  pages 1885--1890 vol.2, Dec 1993.

\bibitem{AutomatedTraffic_2}
B.~{Besselink} and K.~H. {Johansson}.
\newblock String stability and a delay-based spacing policy for vehicle
  platoons subject to disturbances.
\newblock {\em IEEE Transactions on Automatic Control}, 62(9):4376--4391, Sep.
  2017.

\bibitem{gunter2019model_comp}
G.~Gunter.
\newblock Modeling adaptive cruise control vehicles from experimental data:
  Model comparison.
\newblock {\em Proceedings of the IEEE Intelligent Transportation Systems
  Conference}, Oct 2019.

\bibitem{WilsonWard2011}
R.E. Wilson and J.A. Ward.
\newblock Car-following models: fifty years of linear stability analysis – a
  mathematical perspective.
\newblock {\em Transportation Planning and Technology}, 34(1):3--18, 2011.

\bibitem{simon}
D.~Simon.
\newblock {\em Optimal State Estimation}.
\newblock John Wiley \& Sons, Inc., 2006.

\end{thebibliography}

\end{document}